\documentclass{article}
\usepackage{smc2026}
\usepackage[T1]{fontenc}
\usepackage{paralist}% extended list environments
\usepackage[figure,table]{hypcap}% hyperref companion

\usepackage[english]{babel}

\usepackage{todonotes}

\usepackage{multirow}
\usepackage{graphicx}
\usepackage{subcaption}
\usepackage{booktabs}
\usepackage{xcolor}

\newcommand{\etal}{\textit{et al.}}
\newcommand{\ie}{\textit{i.e.} }
\newcommand{\eg}{\textit{e.g.} }

\newcommand{\MusNot}{m}
\newcommand{\Rating}{r}
\newcommand{\GrooveRating}{\Rating_g}
\newcommand{\DanceRating}{\Rating_d}
\newcommand{\ListenRating}{\Rating_l}
\newcommand{\PartyRating}{\Rating_p}

\newcommand{\Embedding}{\phi}

\DeclareMathOperator*{\rtwo}{R^{2}}

\DeclareMathOperator*{\argmin}{arg\,min}

\def\papertitle{Can pre-trained Deep Learning models predict groove ratings?}

\author[1]{\mbox{\firstname{Axel}\lastname{Marmoret}\email{axel.marmoret@imt-atlantique.fr}\orcid{0000-0001-6928-7490}}}
\author[1]{\mbox{\firstname{Nicolas}\lastname{Farrugia}\email{nicolas.farrugia@imt-atlantique.fr}\orcid{0000-0002-1159-3513}}}
\author[2]{\mbox{\firstname{Jan}\middlename{Alexander}\lastname{Stupacher}\email{stupacher@clin.au.dk}\orcid{0000-0002-2179-2508}}}

\affil[1]{\department{IMT Atlantique}\institution{Lab-STICC, UMR CNRS 6285}\city{Brest}\country{France}\affiliationtype{University}}
\affil[2]{\department{Center for Music In the Brain}\institution{Aarhus University}\city{\AA{}rhus}\country{Denmark}\affiliationtype{University}}

\completesetup

\title{\papertitle}
\begin{document}
	\capstartfalse
	\maketitle
	\capstarttrue

	\begin{abstract}
%The abstract should contain about 150-200 words.
This study explores the extent to which deep learning models can predict groove and its related perceptual dimensions directly from audio signals. We critically examine the effectiveness of seven state-of-the-art deep learning models in predicting groove ratings and responses to groove-related queries through the extraction of audio embeddings. Additionally, we compare these predictions with traditional handcrafted audio features. To better understand the underlying mechanics, we extend this methodology to analyze predictions based on source-separated instruments, thereby isolating the contributions of individual musical elements. Our analysis reveals a clear separation of groove characteristics driven by the underlying musical style of the tracks (funk, pop, and rock). These findings indicate that deep audio representations can successfully encode complex, style-dependent groove components that traditional features often miss. Ultimately, this work highlights the capacity of advanced deep learning models to capture the multifaceted concept of groove, demonstrating the strong potential of representation learning to advance predictive Music Information Retrieval methodologies.
\end{abstract}
\section{Introduction}\label{sec:introduction}
%SUPERVISED GROOVE PREDICTION~\cite{chen2025framework}

Groove, an inherently elusive yet fundamentally captivating aspect of music, intrigues researchers and musicians alike due to its powerful ability to evoke movements and emotional responses in listeners. A better understanding of groove not only enriches our comprehension of musical aesthetics, rhythm perception, and social interaction but also holds practical implications for music production, performance, and therapy. This is why the quest to decode groove has led scholars in systematic musicology and cognitive psychology to embark on empirical investigations dissecting its underlying characteristics, mechanisms, and effects.

When describing music with a pronounced and driving rhythm, people often use the noun ``groove'' or the adjective ``groovy''~\cite{stupacher2024text}. These descriptions are sometimes accommodated by expressive gestures such as head bobbing, foot tapping, or other movements associated with this type of music. In the broadest sense, groove can be regarded either as an objective quality of music or a subjective experience. The musical quality of groove is tightly linked to repetitive and layered rhythmic patterns~\cite{zbikowski_modelling_2004}, especially in African-American music genres, such as funk and soul~\cite{camara_groove_2020, iyer_embodied_2002, pressing_black_2002}. The experience of groove is commonly described as a pleasurable urge to move to the music~\cite{madison_experiencing_2006, janata_sensorimotor_2012}. 

Various studies have examined which musical qualities or sonic features best predict the experience of groove. Traditionally, these predictions rely on aesthetic and stylistic interpretations of music, studies of the timing of sounds within and between instrumentalists, or handcrafted audio-signal features. Two of the most common stylistic and rhythmic features linked to groove are syncopation (\ie the placement of sounds on weak metrical positions and the omission of sounds on strong metrical positions) and microtiming (\ie small and expressive timing deviations from the theoretical metrical grid)~\cite{sioros_syncopation_2014, witek_syncopation_2014, keil_theory_1995, senn_effect_2016}. Groove-related audio-signal features include periodicity~\cite{madison_2011}, \textit{cf.}~\cite{stupacher2016audio}, variability of energy and frequency, and energy in bass frequencies~\cite{stupacher2016audio}. Finally, a recent study additionally examined the relationship between the use of groove-related terms in comments on music videos and groove ratings of the same music ~\cite{stupacher2024text}. It was shown that \textit{groove} and \textit{movement} terms were used more often when commenting on songs with higher groove ratings. %This finding suggests that textual representations of music provide an additional basis for groove predictions. 

%In contrast to the aforementioned experimental and musicological approaches, the exploration of rhythm, temporal aspects, and groove in music through computational approaches, particularly music generation with deep learning, has gained substantial momentum in recent years~\cite{moysis2023music}. Such studies rely on the availability of music performance datasets, composed of audio or MIDI data, together with expert annotations (reviewed in~\cite{lerch2021interdisciplinary}). A pivotal advancement in this domain was introduced with the Music Transformer~\cite{huang2018music}, a model that leverages the transformer architecture to generate music with a focus on long-term structure and coherence. This development underscores the potential of deep learning techniques in capturing the intricate temporal dynamics of music, a critical component in the understanding of rhythm and groove. 
%Specifically targeting the concept of groove, Gillick \etal~\cite{gillick2019learning} investigated the generation of groovy drumming sequences using inverse sequence transformations. Their work introduced the GrooveMidi dataset and is a direct contribution to the study of groove in music, demonstrating the efficacy of deep learning models in replicating and understanding the nuances that contribute to the sensation of groove. 
While handcrafted features and stylistic analyses provide valuable insights, they often capture only isolated dimensions of the musical signal. The holistic sensation of groove, however, emerges from complex, non-linear interplays of rhythm, timbre, and micro-dynamics. In recent years, deep learning has emerged as a powerful tool to model such complexities. Within the specific context of groove, computational approaches have successfully tackled rhythm and temporal aspects, primarily through the lens of symbolic data and music generation~\cite{hutchings2017talking, gillick2019learning, makris2022conditional}. For instance, Makris \etal~\cite{makris2022conditional} studied the extent to which drums, generated as accompaniment to guitar and bass, were perceived as "grooving" using listening tests.

Moving beyond symbolic generation, recent work has also begun to explore groove prediction directly from audio. Notably, Chen et al.~\cite{chen2025framework} proposed a framework for groove prediction from audio signals. However, their approach relies strictly on a supervised learning paradigm, meaning its conclusions and feature representations cannot be directly mapped to the capabilities of generalized, unsupervised foundation models. Furthermore, as the field shifts toward massive architectures, a very recent study on Multimodal Large Language Models~\cite{carone2026llms} suggests that such models are currently unable to understand syncopated beats from audio directly, which could largely hinder their abilities to understand groove.
%A very recent study on Multimodal Large Language models~\cite{carone2026llms} suggests that such models are unable to understand syncopated beats from audio directly, which could largely hinder their abilities to understand groove.

These generation-focused, supervised, and multimodal approaches leave an open question: can general-purpose, self-supervised deep neural networks intrinsically understand and predict the subjective experience of groove directly from complex, polyphonic audio?

To bridge this gap, we turn to recent advancements in general-purpose Music Information Retrieval (MIR) models. The MIR landscape has been recently transformed by the rise of many powerful generic deep audio models such as~\cite{huang2022masked, CLAP, niizumi2024m2d, quelennec2025matpac_plus, MERT, zhu2025muq, won2024foundation}. By leveraging Self-Supervised Learning (SSL) paradigms these models bypass the traditional bottleneck of human annotation, learning directly from massive, unlabeled audio corpora. An example of an SSL paradigm is the Masked Acoustic Modeling, where random masks are applied to segments of the input signal (either in the time domain or as spectral features), and the model is optimized to reconstruct the corrupted regions. To succeed in their pre-training tasks, these models must intrinsically capture the underlying structural, timbral, and temporal characteristics of acoustic signals. In addition, multimodal frameworks like CLAP~\cite{CLAP}, M2D~\cite{niizumi2024m2d}, and MuQ~\cite{zhu2025muq} utilize contrastive learning to align these complex audio embeddings with natural language descriptions.

Collectively, these models represent a paradigm shift toward unified musical representations. Because they act as highly effective, frozen feature extractors that treat music as a multi-faceted signal, their high-dimensional embeddings offer a dynamic, automated alternative to traditional handcrafted features. Since the sensation of groove relies heavily on the exact temporal and acoustic hierarchies these models are designed to learn, they are prime candidates for decoding it.

In our study, we aim to harness the potential of these foundational deep audio models to capture groove-related information. We investigate whether state-of-the-art audio embeddings can successfully predict human groove ratings directly from raw audio signals. To this end, we employ an open dataset featuring commercially available songs and their associated groove ratings~\cite{stupacher2024text}. Songs are processed through seven generic deep audio models, and the resulting embeddings are used via linear probing to estimate groove. Furthermore, to better understand the acoustic drivers of groove, we extend our study by applying source separation before embedding extraction. This allows us to isolate and evaluate the extent to which individual instruments (\eg drums, bass) contribute to the overall estimation of groove.

The rest of this paper is organized as follows. In Section~\ref{sec:data}, we describe the dataset in use. Section~\ref{sec:music_processing} is devoted to the deep audio models. Section~\ref{sec:predicting_groove} explains our experimental setups and evaluation. Section~\ref{sec:results_discussion} details the obtained results, and discusses them with respect to prior literature. Finally, Section~\ref{sec:conclusion} summarizes the main conclusions of this paper, and outlines perspectives for future work. 

%%%TODO 
\section{Data and Groove Ratings}
\label{sec:data}
%\subsection{Music and Groove Ratings}

\begin{figure*}[ht]
    \centering
    \includegraphics[width=2\columnwidth]{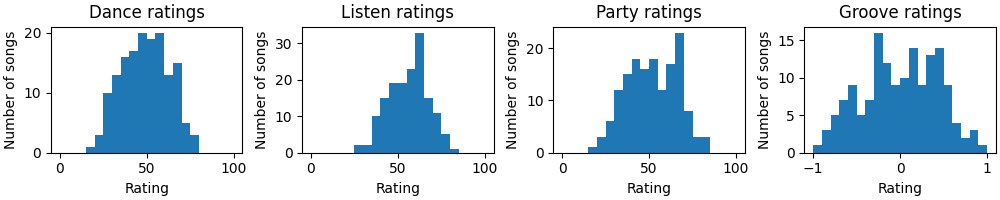}
    \caption{Repartition of the groove ratings for all 148 songs. For each song, results are averaged across all participants.}
    \label{fig:average_ratings}
\end{figure*}

%\axel{TODO (Jan?): expand methodology about groove ratings, as it was requested by reviewers.}

We utilized a dataset comprising 148 songs previously rated on groove in a study by Senn and colleagues~\cite{senn2021taste}. The songs cover a wide range of musical styles, including funk, soul, rap, jazz, pop, disco, rock, and heavy metal. 

To capture the subjective quality of groove, authors of~\cite{senn2021taste} asked listeners to rate each song based on three distinct statements: ``I would like to dance to this music'', ``I like to listen to this music'', and ``This music is great for a party''. Participants had to rate each of these aspects on a scale between 0 (low value) and 100 (high value), thus resulting in three ratings: $\DanceRating(\MusNot)$, $\ListenRating(\MusNot)$, and $\PartyRating(\MusNot)$, respectively for the ``dance'', ``listen'', and ``party'' aspects.

The authors of~\cite{senn2021taste} then derived a ``groove'' rating $\GrooveRating(\MusNot)$ for each song, obtained as the first component of the PCA computed on $\DanceRating(\MusNot)$, $\ListenRating(\MusNot)$, and $\PartyRating(\MusNot)$. Because of the PCA transformation, this final groove rating scales between -1 and 1. The overall distribution of ratings is illustrated in Figure~\ref{fig:average_ratings}, and the complete dataset is openly accessible at \url{https://osf.io/3zjkd}. For more details on how the final scores were obtained and on data collection, readers can refer to the original paper~\cite{senn2021taste}.

% \subsection{YouTube Comments for Text-Based Predictions}
% \label{sec:youtube_comments}
% The textual music representations were taken from a recent study~\cite{stupacher2024text} in which the authors extracted comments on the YouTube videos of the 148 songs that were used in the current study. The resulting 951,998 comments reflect a broad and culturally diverse mix of user-generated content, including reactions, opinions, and feelings~\cite{madden_classification_2013} offering various opportunities for exploratory analyses.

% We additionally used the binary variables defined in~\cite{stupacher2024text}, coding whether comments mentioned one or more words of the following seven groove-related themes: Groove (\textit{e.g.} groovy, grooving), Bonding (\textit{e.g.} together, unite), Event (\textit{e.g.} party, concert), Flow/Smoothness (\textit{e.g.} flow, smooth), Movement (\textit{e.g.} move, dance), Power (\textit{e.g.} energy, push), and Timing (\textit{e.g.} tight, steady).

\section{Audio Signal Processing}
\label{sec:music_processing}
\subsection{Deep Audio Models}
This section provides an overview of the seven general-purpose deep audio models evaluated in this work: AudioMAE~\cite{huang2022masked}, CLAP~\cite{CLAP}, M2D~\cite{niizumi2024m2d}, MATPAC++~\cite{quelennec2025matpac_plus}, MERT~\cite{MERT}, MuQ~\cite{zhu2025muq}, and MusicFM~\cite{won2024foundation}. These models rely on diverse architectures and pre-training objectives. Three of these seven models (MERT~\cite{MERT}, MuQ~\cite{zhu2025muq}, and MusicFM~\cite{won2024foundation}) are exclusively trained on music data. Except for CLAP, all evaluated models were pre-trained using SSL objectives.

\subsubsection{Masked Acoustic Modeling}
Most evaluated models rely on Masked Acoustic Modeling, an SSL paradigm inspired by Masked Language Modeling in Natural Language Processing, where networks are trained to reconstruct a signal from masked or corrupted inputs. Unlike text, where discrete words can be reconstructed directly, audio reconstruction targets are generally modeled indirectly. AudioMAE~\cite{huang2022masked}, for instance, reconstructs missing Mel-spectrogram patches. MERT~\cite{MERT} reconstructs acoustic features (such as CQT and discrete codec tokens) from masked waveforms. MusicFM~\cite{won2024foundation} predicts random projection quantizer targets from masked inputs, while MuQ~\cite{zhu2025muq} predicts discrete Mel Residual Vector Quantization tokens, a standard learned representation in neural audio codecs.

M2D~\cite{niizumi2024m2d} and MATPAC++~\cite{quelennec2025matpac_plus} go a step further in the masking strategy. By splitting the input into complementary visible and masked regions, they then use two models: the first model extracts latent representations from the masked regions, providing targets for the second model to predict based entirely on the visible regions.

\subsubsection{CLAP \& Text-Audio Contrastive Learning}
Unlike the SSL models, CLAP~\cite{CLAP} was trained in a supervised learning fashion. A core strength of CLAP is its text-audio alignment: by using two encoders (one for audio, and one for text) alongside a contrastive learning objective, the model projects both audio and text into a shared latent space. In this study, we extract representations from this multimodal space, utilizing only the audio projection.

M2D~\cite{niizumi2024m2d} and MuQ~\cite{zhu2025muq}, though initially pre-trained using SSL objectives, also undergo a subsequent stage of text-audio multimodal alignment in a setting highly similar to CLAP. For both of these models in our evaluation, we utilize their multimodal projections rather than their purely unimodal audio configurations.

\subsection{MIR features}
As a baseline, we evaluate groove prediction using handcrafted audio features, which have traditionally served as the standard for representing audio signals when predicting subjective musical ratings and behavioral responses. Numerous studies have examined the relationships among musical features, groove ratings, and movement induction. These features include spectral flux (\ie local change in the frequency spectrum) and sub-band flux in low frequency bands~\cite{burger_2012, stupacher_2013, stupacher2016audio}, pulse clarity~\cite{burger_2012}~\textit{cf.}~\cite{stupacher2016audio}, percussiveness~\cite{burger_2012, burger_2013}, event density~\cite{madison_2011}~\textit{cf.}~\cite{stupacher2016audio}, and RMS energy~\cite{stupacher2016audio}. Based on these findings, we use the MIR toolbox~\cite{lartillot2007matlab} to extract the 16 following features from the 148 songs used in the current study: ``mirflux'' including sub-band flux in 10 bands (0-50 Hz, 50-100 Hz, 100-200 Hz, etc.), ``mirpulseclarity'', ``mirpulseclarity \textit{Attack}'', ``mireventdensity'', ``mirrms'', and the standard deviation of the RMS curve extracted with ``mirrms \textit{Frame}''. These so-called ``MIR features'' correspond to the baseline, which we will compare deep embeddings against.

\section{Predicting and Evaluating Groove}
\label{sec:predicting_groove}

\subsection{Open-source statement}
All the experiments carried out in this paper are based on the same open dataset~\cite{stupacher2024text}. We used~\textit{pytorch}~\cite{paszke2019pytorch},~\textit{scikit-learn}~\cite{scikit-learn}, HuggingFace and notably the~\textit{transformers} library~\cite{wolf-etal-2020-transformers} for pretrained deep learning model whenever possible, and MIR toolbox~\cite{lartillot2007matlab}. We release our code for reproducing experiments\footnote{\url{https://github.com/ax-le/deep_groove_prediction}}.

\subsection{Models Formalism}
For a given music $\MusNot$, we compare the audio embeddings of all deep learning models, and the MIR features as a baseline. Embeddings are denoted as $\Embedding(m) \in \mathbb{R}^{e}$, with $e$ the number of dimensions in the embedding space\footnote{Number $e$ of dimensions are: 6144 for AudioMAE, 512 for CLAP, 768 for M2D, 3840 for MATPAC, 1024 for MusicFM, 512 for MuQ, and 9984 for MERT.}. The MIR features are computed with the \textit{MIR} toolbox~\cite{lartillot2007matlab}, as in~\cite{stupacher2016audio}, and 16 features are computed.

The embeddings for AudioMAE, CLAP, MERT, MuQ are computed with \textit{HuggingFace}, respectively using the checkpoints \texttt{hance-ai/audiomae}, \texttt{clap-htsat-unfused}, \texttt{MERT-v1-95M}, and \texttt{MuQ-MuLan-large}. M2D\footnote{\url{https://github.com/nttcslab/m2d}}, MATPAC++\footnote{\url{https://github.com/aurianworld/matpac}}, and MusicFM\footnote{\url{https://github.com/minzwon/musicfm}} (MSD checkpoint) were downloaded from their original repositories.

\subsection{Linear Probing - Ridge Regression}
Given a musical track $\MusNot$ and its corresponding embedding $\Embedding(\MusNot)$, our objective is to accurately predict human-annotated groove ratings. To evaluate the representational power of these embeddings, we employ linear probing via Ridge regression. Ridge regression finds the optimal weight vector $x$ by solving a least-squares problem with an additional $l_2$-norm regularization term, controlled by a penalty parameter $\alpha$, as defined in~\eqnref{RidgeRegression}:
\begin{equation}\label{RidgeRegression}
\underset{x \in \mathbb{R}^{e}}{\argmin} ||\Embedding x - \Rating||_2^2 + \alpha ||x||_2^2
\end{equation}
where $\Embedding \in \mathbb{R}^{n \times e}$ is the embedding matrix for $n$ tracks, and $\Rating$ represents the corresponding human ratings.

We restrict our evaluation to a simple linear model to ensure that performance reflects the representational quality of the embeddings rather than the capacity of a complex predictor. Given our small dataset ($n\approx150$), we implement this using the ``cholesky'' closed-form solver in \textit{scikit-learn}~\cite{scikit-learn} rather than a linear layer in PyTorch. Using a closed-form solver provides a stable, deterministic solution, avoiding the convergence issues and hyperparameter sensitivity associated with training neural networks on very small datasets.

Due to the limited data, we do not perform extensive hyperparameter tuning and fix the regularization coefficient at $\alpha=0.2$ for all models, chosen empirically based on empirical visual inspection.

\begin{table*}[htbp]
\centering
\begin{tabular}{l|cccc}
\toprule
\textbf{Model} & \textbf{$\GrooveRating$} & \textbf{$\DanceRating$} & \textbf{$\ListenRating$} & \textbf{$\PartyRating$} \\ 
\midrule
AudioMAE~\cite{huang2022masked} & $0.23 \pm {\scriptstyle 0.06}$ & $0.29 \pm {\scriptstyle 0.06}$ & $0.07 \pm {\scriptstyle 0.06}$ & $0.21 \pm {\scriptstyle 0.06}$ \\
CLAP~\cite{CLAP} & $0.27 \pm {\scriptstyle 0.08}$ & $0.27 \pm {\scriptstyle 0.06}$ & $\underline{0.22 \pm {\scriptstyle 0.07}}$ & $0.18 \pm {\scriptstyle 0.06}$ \\
M2D~\cite{niizumi2024m2d} & $0.17 \pm {\scriptstyle 0.06}$ & $0.20 \pm {\scriptstyle 0.03}$ & $-0.10 \pm {\scriptstyle 0.07}$ & $0.24 \pm {\scriptstyle 0.07}$ \\
MATPAC++~\cite{quelennec2025matpac_plus} & $\underline{0.35 \pm {\scriptstyle 0.02}}$ & $\underline{0.45 \pm {\scriptstyle 0.01}}$ & $0.07 \pm {\scriptstyle 0.03}$ & $\underline{0.39 \pm {\scriptstyle 0.02}}$ \\
MERT~\cite{MERT} & $0.04 \pm {\scriptstyle 0.07}$ & $0.18 \pm {\scriptstyle 0.06}$ & $-0.12 \pm {\scriptstyle 0.08}$ & $0.05 \pm {\scriptstyle 0.09}$ \\
MuQ~\cite{zhu2025muq} & $\mathbf{0.54 \pm {\scriptstyle 0.03}}$ & $\mathbf{0.59 \pm {\scriptstyle 0.03}}$ & $\mathbf{0.35 \pm {\scriptstyle 0.06}}$ & $\mathbf{0.54 \pm {\scriptstyle 0.02}}$ \\
MusicFM~\cite{won2024foundation} & $0.06 \pm {\scriptstyle 0.07}$ & $0.11 \pm {\scriptstyle 0.07}$ & $-0.10 \pm {\scriptstyle 0.10}$ & $0.08 \pm {\scriptstyle 0.06}$ \\
MIR features~\cite{stupacher2016audio}& $0.03 \pm {\scriptstyle 0.04}$ & $0.05 \pm {\scriptstyle 0.05}$ & $-0.05 \pm {\scriptstyle 0.05}$ & $0.05 \pm {\scriptstyle 0.03}$ \\
\bottomrule
\end{tabular}
\caption{Model Performance Evaluation Results ($\rtwo$ scores)}
\label{tab:r2_audio}
\end{table*}

Finally, we evaluate the models using 4-fold cross-validation. Because performance on a dataset of this size is highly sensitive to the train/test partitioning, we repeat the cross-validation over five independently seeded runs. All reported scores represent the average and standard deviation across these runs to ensure robustness and reproducibility.

\subsection{Model Evaluation with $\rtwo$ score}
We evaluate the relevance of the model using the \textbf{$\rtwo$} score, also called Coefficient of Determination.~The $\rtwo$ score measures the proportion of variance in the original data that is predictable from the embeddings.~The $\rtwo$ score is computed as the ratio of the explained variance by the model to the total variance of the data. Denoting as $\hat{\Rating}$ the averaged rating over all songs, the $\rtwo$ score is given in \eqnref{r_2}.
\begin{equation}\label{r_2}
    1 - \frac{\sum_m (\Embedding(m) x - \Rating(m))^2}{\sum_m (\hat{\Rating} - \Rating(m))^2}
    \end{equation}
An $\rtwo$ score of 0 implies that the model predictions are not better at explaining the variance in the data than a model that simply predicts the mean of the dependent variable, while an $\rtwo$ score of 1 indicates perfect prediction accuracy, meaning that the model predictions perfectly explain the variance in the dependent variable.

\subsection{Analysis on Source-Separated Stems}
In order to evaluate the contribution of each instrument to the groove ratings, we also estimate groove ratings based on isolated instruments. To this end, we perform source separation using the State-of-the-Art deep learning model Hybrid-Demucs~\cite{rouard2023hybrid}, leading to the 4-stem separation in bass, drums, vocals, and ``other''. We perform the same steps with source-separated stems as with the full tracks, namely, we extract deep audio embeddings, followed by ridge regression to predict groove ratings. We evaluate results using $\rtwo$.

\subsection{Visualizations}
 In order to bring qualitative insights on model predictions,  we use two types of visualizations. First, we use scatter plots to visualize the quality of the linear relationship between predicted values (averaged across cross-validation runs) and groove ratings, considered as ground truth. In addition, we plot the two principal components of embeddings extracted from full audio and separated stems, and interpret those jointly with groove ratings. 
 
\section{Results and Discussion}
\label{sec:results_discussion}
\subsection{Deep Audio Model Comparison in Groove Prediction}
\begin{figure*}[htb!]
     \centering
    \begin{minipage}{0.9\textwidth} % Change 0.9 to 0.8 or 0.7 to shrink everything
     % --- Row 1 ---
     \begin{subfigure}[b]{0.48\textwidth}
         \centering
         \includegraphics[width=\textwidth]{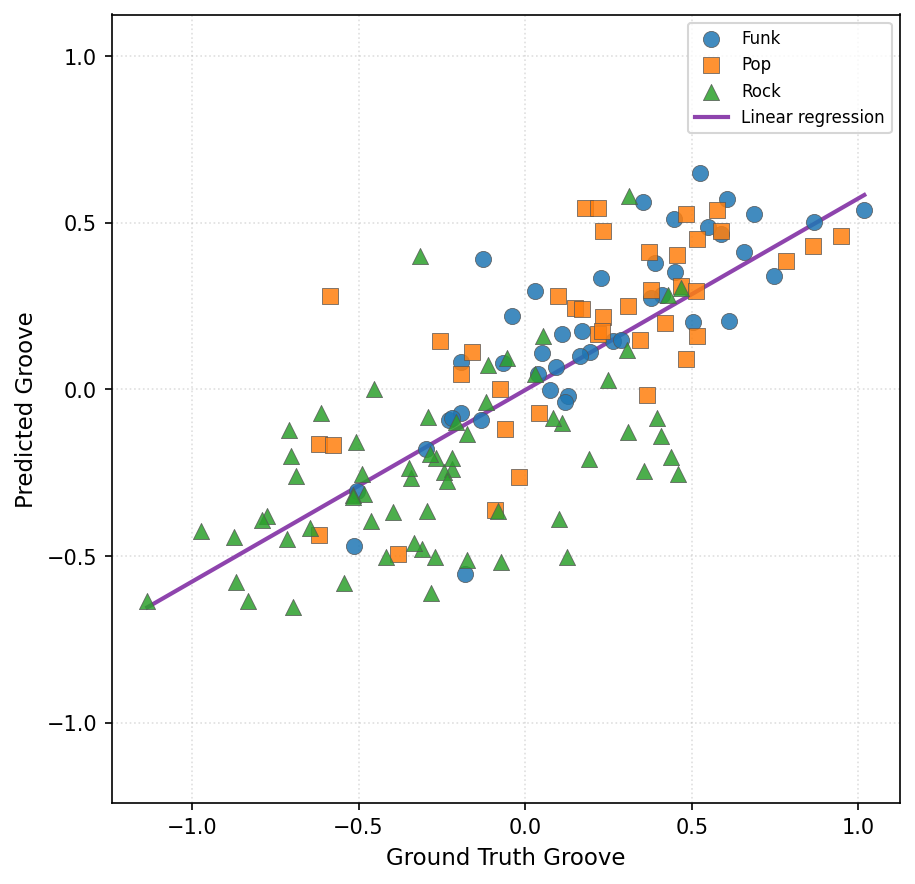}
         \caption{Groove rating}
         \label{fig:scatter_groove_muq}
     \end{subfigure}
     \hfill
     \begin{subfigure}[b]{0.48\textwidth}
         \centering
         \includegraphics[width=\textwidth]{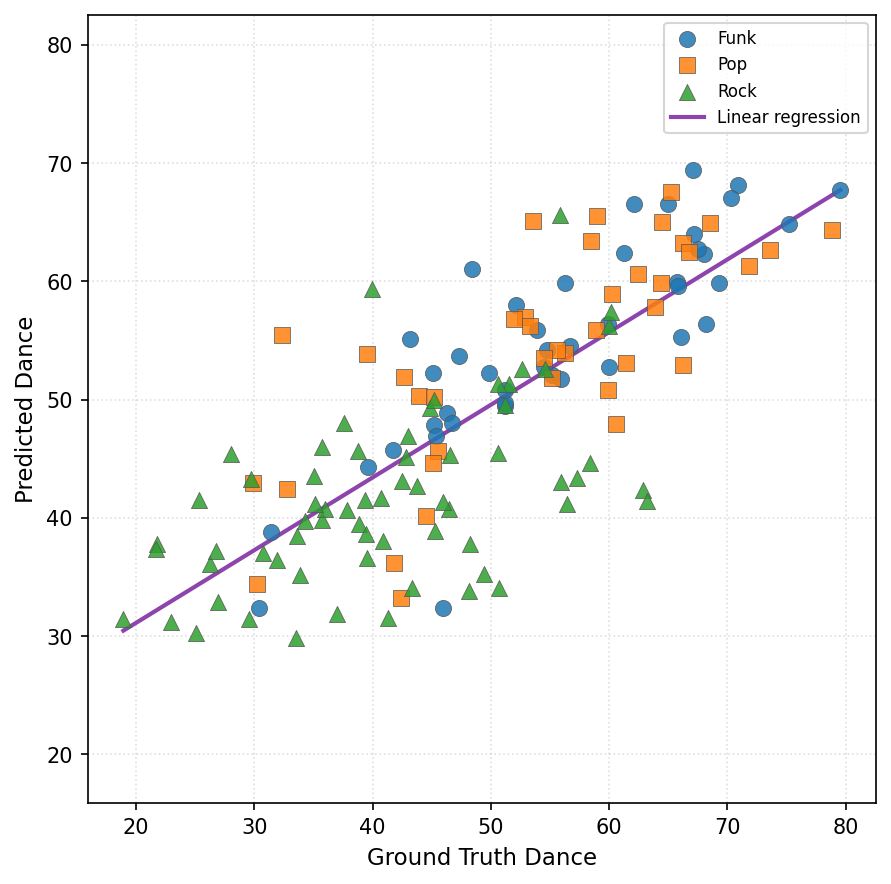}
         \caption{Dance rating}
         \label{fig:scatter_dance_muq}
     \end{subfigure}

     %\vspace{1em}

     % --- Row 2 ---
     \begin{subfigure}[b]{0.48\textwidth}
         \centering
         \includegraphics[width=\textwidth]{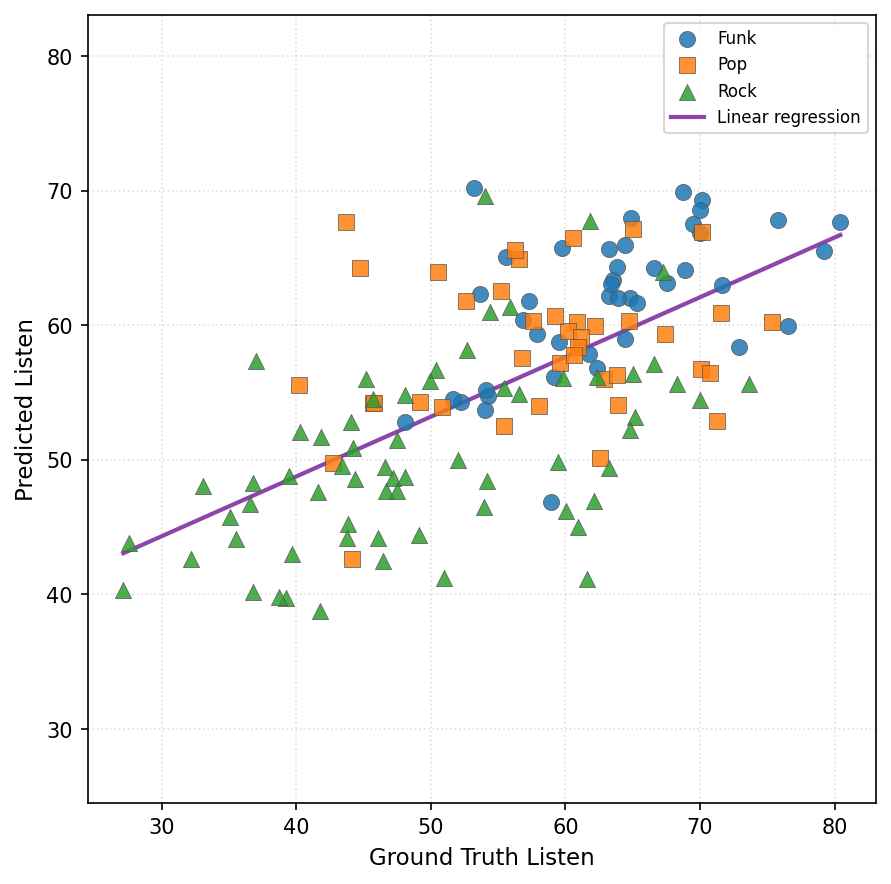}
         \caption{Listen rating}
         \label{fig:scatter_listen_muq}
     \end{subfigure}
     \hfill
     \begin{subfigure}[b]{0.48\textwidth}
         \centering
         \includegraphics[width=\textwidth]{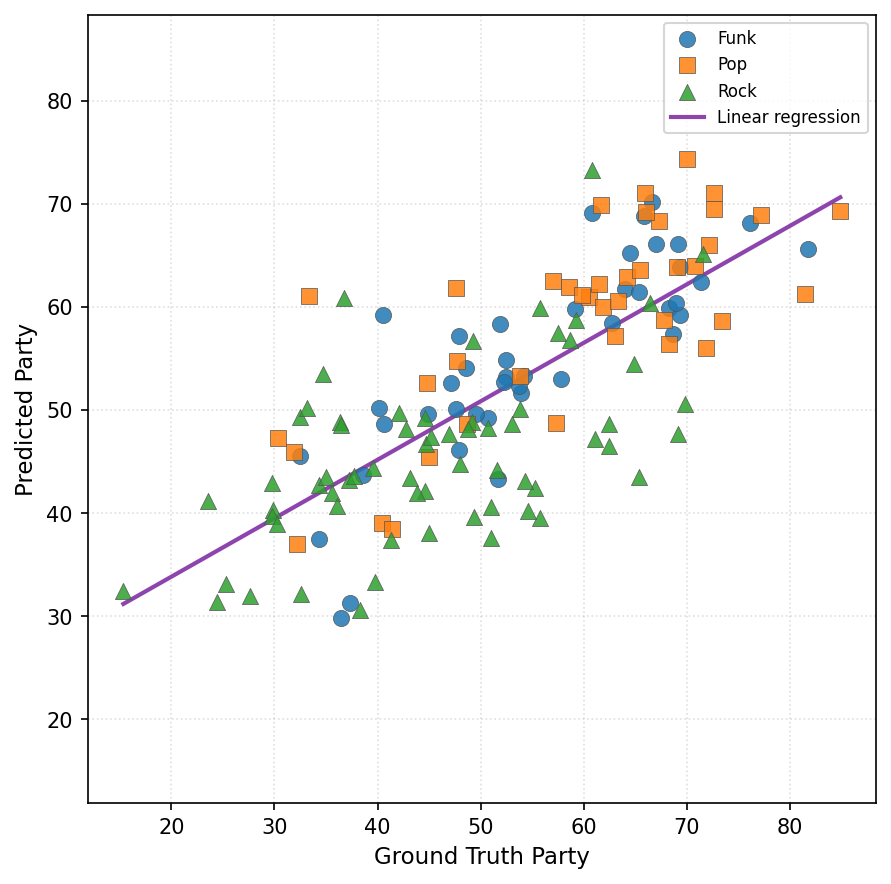}
         \caption{Party rating}
         \label{fig:scatter_party_muq}
     \end{subfigure}

    \end{minipage}
     \caption{Scatter plots of the predicted values against the ground truth values, for all ratings, and for the MuQ embeddings.}
     \label{fig:scatter_all_ratings}
\end{figure*}

The MuQ model significantly outperforms the other models and the traditional MIR feature extraction approach across all metrics (\tabref{tab:r2_audio}). When predicting groove ratings, the MuQ model achieves an $\rtwo$ score of $0.54$, which demonstrates a moderate to strong ability to explain the variance in groove ratings. This pattern of results is consistent across answers to the questions related to ``dance'', ``listen'', and ``party''. Scatter plots of the prediction of MuQ ratings compared with the ground truth are presented in Figure~\ref{fig:scatter_all_ratings}.

Overall, these results demonstrate that deep audio embeddings can capture groove-related information from audio signals, despite the absence of explicit training on such information. On the contrary, the MIR features we extracted cannot reliably predict groove ratings, as demonstrated by the very low (or negative) $\rtwo$ scores obtained. 

Results in Table~\ref{tab:r2_audio} reveal that all evaluated models achieve their highest predictive accuracy on responses to the ``dance'' question. On the other hand, the lowest levels of predictive accuracy for all models are obtained for the ``like to listen'' question. These results suggest that the attribute of danceability represents the most quantifiable and predictable element of a song's groove characteristics. Based on the common definition of the groove experience as a pleasurable urge to move to music (\eg~\cite{janata_sensorimotor_2012,stupacher_sweet_2022}), ``dance'' ratings primarily reflect the motor component, whereas ``listen'' ratings mostly capture the pleasure component of groove. Our findings therefore suggest that the pleasure component of groove is more subjective and heterogeneous than the motor component. 

This discrepancy might be further explained by the social nature of the ratings: while ``listening'' is often a private, personal experience primarily used for mood regulation and self-awareness~\cite{schafer2013psychological}, ``dancing'' and ``partying'' are inherently social activities that facilitate interpersonal synchrony and social bonding~\cite{tarr2014music}. This social context may foster a stronger group consensus (or intersubjective agreement) on what makes a song danceable, creating a stable and predictable signal~\cite{solberg2017pleasurable}, whereas individual listening preferences remain much more varied and dependent on personal psychosocial factors~\cite{rentfrow2011structure}.

\subsection{Predicting Groove from Source Separated Stems}

\begin{table*}[htb!]
\centering
\begin{tabular}{lcccc}
\toprule
\textbf{Instrument (isolated)} & \textbf{$\GrooveRating$} & \textbf{$\DanceRating$} & \textbf{$\ListenRating$} & \textbf{$\PartyRating$} \\ 
\midrule
Vocals          & $0.15 \pm {\scriptstyle 0.04}$ & $0.20 \pm {\scriptstyle 0.04}$ & $0.04 \pm {\scriptstyle 0.03}$ & $0.13 \pm {\scriptstyle 0.04}$ \\
Bass            & $\underline{0.29 \pm {\scriptstyle 0.02}}$ & $\underline{0.32 \pm {\scriptstyle 0.02}}$ & $\underline{0.27 \pm {\scriptstyle 0.03}}$ & $0.22 \pm {\scriptstyle 0.01}$ \\
Drums           & $0.25 \pm {\scriptstyle 0.03}$ & $0.27 \pm {\scriptstyle 0.02}$ & $0.12 \pm {\scriptstyle 0.06}$ & $\underline{0.24 \pm {\scriptstyle 0.02}}$ \\
Other           & $\mathbf{0.42 \pm {\scriptstyle 0.02}}$ & $\mathbf{0.47 \pm {\scriptstyle 0.01}}$ & $\mathbf{0.33 \pm {\scriptstyle 0.04}}$ & $\mathbf{0.34 \pm {\scriptstyle 0.03}}$ \\
%Drums + Bass   & $\underline{0.31 \pm {\scriptstyle 0.05}}$ & $\underline{0.35 \pm {\scriptstyle 0.04}}$ & $0.15 \pm {\scriptstyle 0.05}$ & $\underline{0.30 \pm {\scriptstyle 0.05}}$ \\
\midrule
Full audio      & $0.54 \pm {\scriptstyle 0.03}$ & $0.59 \pm {\scriptstyle 0.03}$ & $0.35 \pm {\scriptstyle 0.06}$ & $0.54 \pm {\scriptstyle 0.02}$ \\
\bottomrule
\end{tabular}
\caption{Groove Prediction results $\left(\rtwo \right)$ with MuQ embeddings computed on isolated instruments.}
\label{tab:source_separation_results}
\end{table*}

In addition, as shown in~\tabref{tab:source_separation_results}, model performance is significantly higher when using the ``other'' stem (comprising harmonic and melodic instruments) compared to using the drums or bass alone. Specifically, the ``other'' stem achieved an average $\rtwo$ score for groove of 0.42, whereas the drums and bass reached only 0.25 and 0.29 on average, respectively. This disparity suggests that the features most predictive of groove in our models are not found in the primary rhythmic section, but are instead concentrated in the harmonic and instrumental layers.

% \begin{figure*}[htbp]
%      \centering
%     %\begin{minipage}{0.9\textwidth} % Change 0.9 to 0.8 or 0.7 to shrink everything
%      % --- Row 1 ---
%      \begin{subfigure}[b]{0.45\textwidth}
%          \centering
%          \includegraphics[width=\textwidth]{figs/PCA/vocals.png}
%          \vspace{-0.5cm}
%          \caption{Vocals}
%          \label{fig:pca_vocals}
%      \end{subfigure}
%      \hfill
%      \begin{subfigure}[b]{0.45\textwidth}
%          \centering
%          \includegraphics[width=\textwidth]{figs/PCA/bass.png}
%          \vspace{-0.5cm}
%          \caption{Bass}
%          \label{fig:pca_bass}
%      \end{subfigure}

%      %\vspace{1em}

%      % --- Row 2 ---
%      \begin{subfigure}[b]{0.45\textwidth}
%          \centering
%          \includegraphics[width=\textwidth]{figs/PCA/drums.png}
%          \vspace{-0.5cm}
%          \caption{Drums}
%          \label{fig:pca_drums}
%      \end{subfigure}
%      \hfill
%      \begin{subfigure}[b]{0.45\textwidth}
%          \centering
%          \includegraphics[width=\textwidth]{figs/PCA/other.png}
%          \vspace{-0.5cm}
%          \caption{``Other'' instruments}
%          \label{fig:pca_instr}
%      \end{subfigure}

%     %\end{minipage}

%      \caption{Visualization of the projection on the first two principal components (PCA) of audio embeddings from the OpenMuQ model derived from source-separated instrumental stems.}
%      \label{fig:pca_all_embed}
% \end{figure*}

\begin{figure*}[htb!]
    \centering
    \includegraphics[width=\textwidth]{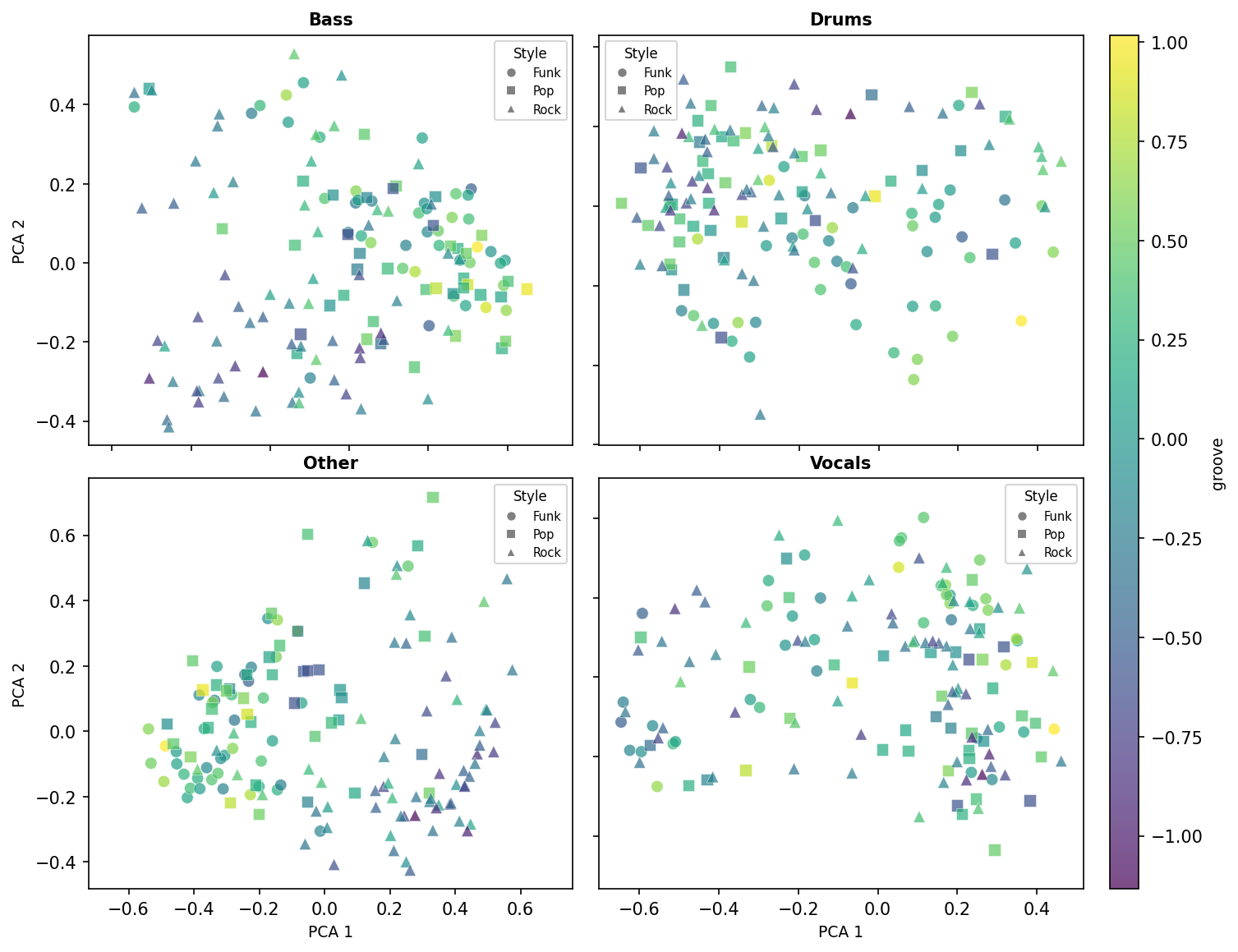}
    \caption{Visualization of the projection on the first two principal components (PCA) of audio embeddings from the MuQ model derived from source-separated instrumental stems}
    \label{fig:pca_all_embed}
\end{figure*}

PCA visualizations, presented in Figure~\ref{fig:pca_all_embed}, provide a structural explanation for this finding, indicating that the model likely utilizes musical genre as a dominant proxy for groove. The ``other'' stem embeddings exhibit a pronounced separation of tracks by musical style, with Funk, Pop, and Rock forming almost distinct clusters. This separation is noticeably weaker and more overlapping in projections based on bass, drum, or vocal stems. A qualitative evaluation of the source-separated stems supports this stylistic differentiation: while Rock tracks feature relatively homogeneous ``other'' content (often just distorted electric guitars) and Pop tracks show constrained diversity, Funk tracks (which carry the highest average groove ratings) frequently include a richer, heterogeneous mix of instruments, notably involving prominent horn sections. This result aligns with the common view that groove is related to musical style~\cite{stupacher2024text}. Hence, we hypothesize that the MuQ model successfully predicts groove by leveraging style-dependent spectral cues, such as the distinct signatures of funk horn sections.%, providing empirical weight to the view that groove is deeply tied to musical genre.

\section{Conclusion and Perspectives}
\label{sec:conclusion}

In this paper, we have shown that groove ratings can be predicted using embeddings extracted from deep audio models. The MuQ model significantly outperforms traditional MIR features and other deep embeddings across all metrics. Predictive accuracy is notably higher for ``dance'' than for ``listen'' ratings, supporting the hypothesis that the motor component of groove is more objectively quantifiable and driven by social consensus than the subjective, heterogeneous pleasure of listening. Furthermore, source-separation analysis reveals that melodic and harmonic ``other'' stems are more predictive than drums or bass, suggesting the model captures groove by leveraging genre-specific spectral signatures (such as funk horn sections) rather than relying solely on traditional rhythmic foundations.
Future work will employ interpretability and visualization techniques to identify the specific features, such as microtiming~\cite{senn2021taste}, that underpin the model’s performance. Additionally, we plan to utilize multimodal models like MotionBeat~\cite{wang2026motionbeat} to correlate audio embeddings with physical motion-capture data, further quantifying the motor component of groove. 

Ultimately, this study demonstrates how machine learning can bridge the gap between computational signal processing and the subjective human experience of music, movement, and emotion.

%Overall, we demonstrate that it is possible to predict groove ratings with audio signals using deep learning models. Prediction scores of all deep audio models are higher than tested handcrafted features, with some obtaining way higher scores than others (MuQ in particular). 

%Future work will attempt a better characterization of important features related to groove, using visualization and interpretability techniques. For instance, some studies (\eg~\cite{senn2021taste}) have correlated different acoustic features (such as microtiming) with groove, which could be worth studying in our context. 

%We could also relate the audio acoustic features with captured motion of people using multimodal models such as~\cite{wang2026motionbeat}.

%This study opens up opportunities for further research into the relationship between music, movement, and emotion. It shows how machine learning can help connect computational techniques with our understanding of musical experiences.
	
	%%%%%%%%%%%%%%%%%%%%%%%%%%%%%%%%%%%%%%%%%%%%%%%%%%%%%%%%%%%%%%%%%%%%%%%%%%%%%
	%bibliography here
	\bibliography{smc2026bib}
	
\end{document}